\documentclass[twocolumn,prb,showpacs,preprintnumbers,superscriptaddress,amsmath,amssymb,10pt,aps,floatfix]{revtex4-1}
\usepackage{graphicx}
\usepackage{dcolumn}
\usepackage{color}
\usepackage{bm}
\usepackage[hidelinks]{hyperref}
\hypersetup{
    colorlinks,
    citecolor=blue,
    filecolor=blue,
    linkcolor=blue,
    urlcolor=blue
}

\usepackage{graphicx,rotating,subfigure,amsmath,amsfonts,amssymb,delarray}
\usepackage{dsfont}
\usepackage[T1]{fontenc}
\numberwithin{equation}{section}

\newcommand{\im}{\textrm{i}}

\DeclareMathOperator{\e}{e}

\DeclareMathOperator{\Tr}{Tr}

\allowdisplaybreaks

\begin{document}

\title{The fate of dynamical phase transitions at finite temperatures and in open systems}
\author{N.~Sedlmayr}
\email{ndsedlmayr@gmail.com}
\affiliation{Department of Physics and Medical Engineering, Rzesz\'ow University of Technology, al.~Powsta\'nc\'ow Warszawy 6, 35-959 Rzesz\'ow, Poland}
\author{M.~Fleischhauer}
\affiliation{Department of Physics and Research Center OPTIMAS, University of Kaiserslautern, Germany}
\author{J.~Sirker}
\affiliation{Department of Physics and Astronomy, University of Manitoba, Winnipeg R3T 2N2, Canada}

\date{\today}

\begin{abstract}
When a quantum system is quenched from its ground state, the time
evolution can lead to non-analytic behavior in the return rate at
critical times $t_c$. Such {\it dynamical phase transitions} (DPT's)
can occur, in particular, for quenches between phases with different
topological properties in Gaussian models. In this paper we discuss
Loschmidt echos generalized to density matrices and obtain results for
quenches in closed Gaussian models at finite temperatures as well as
for open system dynamics described by a Lindblad master
equation. While cusps in the return rate are always smoothed out by
finite temperatures we show that dissipative dynamics can be
fine-tuned such that DPT's persist.
\end{abstract}

\maketitle

\section{Introduction}
The macroscopic properties of a quantum system in equilibrium can be
understood from the appropriate thermodynamic potential. Studies of
Lee-Yang zeros of the grand-canonical potential as a function of a
complex fugacity or of Fisher zeros of the canonical potential as a
function of complex temperature, in particular, have significantly
contributed to our understanding of equilibrium phase
transitions.\cite{Yang1952,Lee1952,Fisher1965,Bena2005} In recent
years, there have been attempts to follow a similar approach to
non-equilibrium dynamics. For quench dynamics in closed quantum
systems it has been suggested that {\it dynamical phase transitions}
(DPT's) can be defined based on the Loschmidt echo\cite{Heyl2013}
\begin{equation}
\label{Lo}
{\cal L}_0(t)=\langle\Psi_0|\e^{-iH_1t}|\Psi_0\rangle\,.
\end{equation}
Here $|\Psi_0\rangle$ is the pure quantum state before the quench and
$H_1$ the time-independent Hamiltonian responsible for the unitary
time evolution. The Loschmidt echo has the form of a partition
function with boundaries fixed by the initial state. In analogy to the
Fisher zeros in equilibrium one can thus study the zeros of the
Loschmidt echo for complex time $t$. In Ref.~\onlinecite{Heyl2013} it
has been shown that for the specific case of the transverse Ising
model these zeros form lines in the complex plane which cross the real
axis only for a quench across the equilibrium critical point.

In a many-body system one expects that the overlap between the
time-evolved and the initial state is in general exponentially small
in system size in analogy to the {\it Anderson orthogonality catastrophe}
in equilibrium.\cite{Anderson1967} To obtain a non-zero and well-defined
quantity in the thermodynamic limit it is thus useful to consider
the return rate
\begin{equation}
\label{return}
l_0(t)=-\lim_{L\to\infty}\frac{1}{L}\ln|{\cal L}_0(t)| \,.
\end{equation}
where $L$ is the system size. Zeros in ${\cal L}_0(t)$ at critical
times $t_c$ then correspond to non-analyticities (cusps or
divergencies) in
$l_0(t)$.\cite{Heyl2013,Karrasch2013,Andraschko2014,Halimeh2017,Homrighausen2017,Jafari2017a}
It is, however, important to stress that in contrast to the
particularly simple case of the transverse Ising model there is in
general no one-to-one correspondence between dynamical and equilibrium
phase transitions.\cite{Andraschko2014,Vajna2014} It is possible to
find non-analytical behavior of the return rate without crossing an
equilibrium critical point in the quench, and one can cross a critical
line without non-analyticities in $l_0(t)$ being present.  For
one-dimensional topological systems it has been shown, in particular,
that crossing a topological phase transition in the quench always
leads to a DPT but the opposite does not have to be
true.\cite{Vajna2015} Thus there are still some issues about the
appropriateness of the Loschmidt echo as a useful indicator.
Nevertheless the notion of a dynamical phase transition is an exciting
concept extending key elements of many-body physics to
non-equilibrium.

Lately, DPT's have also been studied experimentally. In
Ref.~\onlinecite{Flaschner2016} vortices in a gas of ultracold
fermions in an optical lattice were studied and their number
interpreted as a dynamical order parameter which changes at a
DPT. Even more closely related to the described formalism to classify
DPT's is an experiment where a long-range transverse Ising model was
realized with trapped ions. In this case the time-evolved state was
projected onto the initial state which allowed access to the Loschmidt
echo \eqref{Lo} directly.\cite{Jurcevic2017}

While these experiments are an exciting first step to test these
far-from-equilibrium theoretical concepts they also lead to a number
of new questions. Chief among them is the question how experimental
imperfections affect the Loschmidt echo and DPT's. On the one hand,
the initial state is typically not a pure but rather a mixed state at
a certain temperature $T$. This raises the question how the Loschmidt
echo can be generalized to thermal states. On the other hand, the
dynamics is also typically not purely unitary. Decoherence and
particle loss processes do affect the dynamics as well, requiring a
generalization of \eqref{Lo} to density matrices. Finally dynamical processes
and phase transitions can be induced entirely by coupling to reservoirs in which
case no pure-state or $T=0$ limit exists.\cite{HoeningMoos}

In this paper we will address these questions. In Sec.~\ref{Sec_Lo} we
discuss various different ways to generalize the Loschmidt echo to
finite temperatures. We concentrate, in particular, on projective
measurements of time-evolved density matrices relevant for example,
for trapped ion experiments, as well as on a proper distance measure
between the initial and the time-evolved density matrix following
Refs.~\onlinecite{Zanardi2007a,CamposVenuti2011}. We study both of
these generalized Loschmidt echos for the case of unitary dynamics of
Gaussian fermionic models in Sec.~\ref{Gaussian}. As examples, we
present results for the transverse Ising and for the
Su-Schrieffer-Heeger (SSH) model. In Sec.~\ref{Open} we consider the
generalized Loschmidt echo for open-system dynamics of Gaussian
fermionic models described by a Lindblad master equation (LME). A
short summary and conclusions are presented in Sec.~\ref{Concl}.

\section{The Loschmidt echo}
\label{Sec_Lo}

We will first review some properties of the standard Loschmidt echo
for unitary dynamics of pure states in Sec.~\ref{zero_Lo} before
discussing several possible generalizations to mixed states in
Sec.~\ref{DM_Lo}.

\subsection{Pure states}
\label{zero_Lo}

The Loschmidt echo for unitary dynamics of a pure state is defined by
Eq.~\eqref{Lo}. Its absolute value can be used to define a metric in
Hilbert space $\phi=\arccos|{\cal L}_0(t)|$ with $0\leq |{\cal
L}_0(t)| \leq 1$ which characterizes the distance between the initial
state $|\Psi_0\rangle$ and the time-evolved state $|\Psi(t)\rangle
=\e^{-iH_1t}|\Psi_0\rangle $.\cite{Nielsen_book} From this point of view the Loschmidt
echo is a time-dependent version of the {\it fidelity}
$F=|\langle\Psi_0|\Psi_1\rangle|$ which has been widely used to study
equilibrium phase
transitions.\cite{Vidal2003a,Venuti2007,Schwandt2009,Zhou2008,Zanardi2006,Zanardi2007a,Zanardi2007b,Chen2007,You2007,Yang2007,Dillenschneider2008,Sarandy2009,Sirker2010,Sirker2014a,Koenig2016}
Because of the Anderson orthogonality catastrophy one has to consider
a fidelity density $f=-\lim_{L\to\infty}\ln|F|/L$ for a many-body
system in the thermodynamic limit $L\to \infty$ in analogy to the
Loschmidt return rate defined in Eq.~\eqref{return}. If
$|\Psi_0\rangle$ and $|\Psi_1\rangle$ are both ground states of a
Hamiltonian $H(\lambda)$ for different parameters $\lambda$ then the
fidelity susceptibility $\chi_f=(\partial^2
f)/(\partial\lambda)^2|_{\lambda=\lambda_c}$ will typically diverge at
an equilibrium phase transition. Similarly, one might expect that a
quench can lead to states $|\Psi(t_c)\rangle$ at critical times $t_c$
which are orthogonal to the initial state implying ${\cal L}_0(t_c)=0$
and resulting in a non-analyticity in the return rate $l_0(t_c)$. A
peculiarity of the return rate is that its non-analyticity does not
only depend on the properties of the initial and final Hamiltonian
before and after the quench but also on time. For a quench from $H_0$
to $H_1$, in particular, the critical time $t_c$ will in general
depend upon if one starts with the ground state of the initial
Hamiltonian or some excited eigenstate.

\subsection{Mixed states}
\label{DM_Lo}
\subsubsection{Loschmidt echo as a metric}
If the Loschmidt echo is primarily seen as defining a metric in
Hilbert space, then it is natural to ask if a similar metric can also
be defined for density matrices $\rho(t)$. In order for the
generalized Loschmidt echo $|{\cal L}_\rho(\rho(0),\rho(t))|$ to give
rise to a proper measure of distance in the space of density matrices
we want the following relations to hold
\begin{itemize}
\item[1)] $0\leq |{\cal L}_\rho(\rho(0),\rho(t))|\leq1$ and $|{\cal L}_\rho(\rho(0),\rho(0))|=1$,
\item[2)] $|{\cal L}_\rho(\rho(0),\rho(t))|=1$ iff $\rho(0)=\rho(t)$, and
\item[3)] $|{\cal L}_\rho(\rho(0),\rho(t))|=|{\cal L}_\rho(\rho(t),\rho(0))|$.
\end{itemize}
Without time dependence, this problem reduces again to the definition
of a fidelity for density matrices.\cite{Bures1969,Uhlmann1976,Jozsa1994} A
direct generalization of this fidelity leads to\cite{Zanardi2007a,CamposVenuti2011}
\begin{equation}
\label{LoT}
{\cal L}_\rho(t)\equiv |{\cal L}_\rho(\rho(0),\rho(t))|=\Tr\sqrt{\sqrt{\rho(0)}\rho(t)\sqrt{\rho(0)}}\,.
\end{equation}
Note that this definition satisfies $\lim_{\beta\to\infty}{\cal
L}_\rho(t)=|{\cal L}_0(t)|$ if $\rho(0)$ is a thermal density matrix
and the time evolution is unitary. $\beta=T^{-1}$ is the inverse
temperature with $k_B=1$. ${\cal L}_\rho(t)$ is symmetric between
$\rho(0)$ and $\rho(t)$ and also satisfies the other conditions
above. The induced metric $\phi=\arccos[{\cal L}(t)]$ also fulfills
the triangle inequality.\cite{Nielsen_book} From this point of view,
Eq.~\eqref{LoT} is thus the proper generalization of the Loschmidt
echo to density matrices. Despite its relatively complicated
appearance, $|{\cal L}_\rho(\rho_1,\rho_2)|$ has a straightforward
physical meaning.\cite{Jozsa1994} If we understand $\rho_1$ and
$\rho_2$ as reduced density matrices obtained by a partial trace over
a larger system which is in a pure state $|\phi_{1,2}\rangle$
respectively, then $|{\cal L}_\rho(\rho_1,\rho_2)|=\max
|\langle\phi_1|\phi_2\rangle|$ where the maximum is taken over all
purifications of $\rho_1$ and $\rho_2$, respectively. I.e., ${\cal
L}_\rho$ provides the purification to the states in the enlarged
Hilbert space which are as parallel as possible and consistent with
the mixed states of the subsystem.

A seemingly simpler and more straightforward generalization such as  
\begin{equation}
\label{Ltilde}
|\tilde {\cal L}_\rho(t)|=\sqrt{\frac{\Tr\{\rho(0)\rho(t)\}}{\Tr\rho^2(0)}}
\end{equation}
does, in general, not fulfill the conditions above. If we start, for
example, in a completely mixed state $\rho(0)=\sum_n
\frac{1}{N}|\Psi_n\rangle\langle \Psi_n|$ and evolve under dissipative dynamics to a pure state
$\rho(t\to\infty)=|\Psi_0\rangle\langle \Psi_0|$ then $|\tilde {\cal
L}_\rho(0)|=|\tilde {\cal L}_\rho(\infty)|=1$ which clearly is also
not a desirable property. Using a spectral representation in a basis
where $\rho(0)=\sum_n p_n |\Psi^0_n\rangle\langle \Psi^0_n| $ is
diagonal, Eq.~\eqref{Ltilde} for the special case of unitary time
evolution can be represented as
\begin{equation}
\label{Ltilde_spec}
|\tilde {\cal L}_\rho(t)|^2={\frac{\sum_{m,n} p_mp_n |\langle \Psi^0_m|\e^{-iHt}|\Psi_n^0\rangle|^2}{\sum_n p_n^2}} \, ,
\end{equation}
where $p_n$ are weights with $\sum_n p_n=1$.

In Sec.~\ref{Gaussian} we will investigate ${\cal L}_\rho(t)$ for unitary
dynamics in Gaussian models with $\rho(0)$ being a canonical density
matrix at a given finite temperature $T$. At the same time, we will
also briefly discuss the result for $\tilde {\cal L}_\rho(t)$ which---for
unitary dynamics---in this specific case does fulfill $0\leq |\tilde
{\cal L}_\rho(t)|\leq 1$. This is no longer the case for open system dynamics
described by an LME and we will therefore exclusively discuss
${\cal L}_\rho(t)$ in Sec.~\ref{Open}.

\subsubsection{Projection onto a pure state}
While \eqref{LoT} allows to generalize the properties of the Loschmidt
echo as a metric to density matrices, ${\cal L}_\rho(t)$ might not
necessarily be the quantity measured experimentally. In
Ref.~\onlinecite{Jurcevic2017}, for example, DPT's in the transverse
Ising model have been investigated using a system of trapped ions. In
this experiment the system is prepared in an initial configuration,
the system is then time evolved and the Loschmidt echo measured by a
projection. If the system is prepared in a pure state and the
projection is onto the same pure state then the Loschmidt echo
\eqref{Lo} is measured. Here we want to consider the case that the
preparation of the system is not ideal---leading to a mixed instead of
a pure state---while the projection is still onto the ground state of
the initial Hamiltonian. I.e., we consider the case that only one of
the states is impure. In this case we can define a generalized
Loschmidt echo by replacing $\rho(0)\to |\Psi_0^0\rangle\langle
\Psi_0^0|$ in Eq.~\eqref{LoT} leading to 
\begin{eqnarray}
\label{Lproj}
|{\cal L}_p(t)|^2 &=& {\langle
\Psi_0^0|\rho(t)|\Psi_0^0\rangle}/{\langle\Psi_0^0|\rho(0)|\Psi_0^0\rangle}
\\ &=& \sum_n \frac{p_n}{p_0} |\langle \Psi_0^0
|\e^{-iHt}|\Psi_n^0\rangle|^2 \nonumber \, .
\end{eqnarray}
The second line is a spectral representation in the eigenbasis
of $\rho(0)$ and we have introduced a normalization factor such that
${\cal L}_p(0)=1$. Note that for a thermal initial density matrix
$\lim_{\beta\to\infty}|{\cal L}_p(t)|^2 = |{\cal L}_0(t)|^2$. In
Sec.~\ref{Gaussian} we will also investigate this generalization of
the Loschmidt echo for unitary dynamics and present results for
experimentally relevant cases such as the transverse Ising and the SSH
model.

\subsubsection{Alternative generalizations}
The definition of a generalized Loschmidt echo for mixed states is not
unique and several other possible generalizations have been discussed
previously in the literature. In Ref.~\onlinecite{Dutta} and
Ref.~\onlinecite{Heyl2017} the quantity
\begin{eqnarray}
\label{Lav}
{\cal L}_{\textrm{av}} &=& \Tr\left\{\rho(0)U(t)\right\} \\
&=& \sum_n p_n \langle \Psi^0_n|\e^{-iH_1 t}|\Psi^0_n\rangle \nonumber
\end{eqnarray}
is considered where $U(t)$ is the time-evolution operator. From the
spectral representation for unitary time evolution with a
time-independent Hamiltonian shown in the second line of
Eq.~\eqref{Lav} it is clear that this generalization measures an
average over pure-state Loschmidt echos rather than the `overlap'
between mixed states as defined in Eq.~\eqref{LoT}. Also in contrast
to \eqref{Ltilde_spec} only diagonal terms enter; Eq.\eqref{Lav}
cannot be used to define a measure of distance between {\it two}
density matrices. For a generic Gibbs ensemble one expects, in
general, that ${\cal L}_{\textrm{av}}=0$ is only possible if $p_0=1$,
since even if the Loschmidt echos of different states $\vert
\Psi_n^0\rangle$ will vanish at some time, the corresponding critcial
times will in general be different.  For a Gaussian model in a {\it
generalized Gibbs ensemble}, where the occupation of each $k$-mode is
individually conserved, zeros are however also possible at finite
temperatures.\cite{Heyl2017}

A similar approach---motivated by the characteristic function of
work\cite{Talkner2007}---was also used in
Ref.~\onlinecite{Abeling2016} where the specific case of a canonical
density matrix as initial condition was considered and a generalized
Loschmidt echo defined by
\begin{eqnarray}
\label{tildeLav}
\tilde {\cal L}_{\textrm{av}} &=& \frac{1}{Z}\Tr\left\{\e^{iH_1t}\e^{-iH_0t}\e^{-\beta H_0}\right\} \\
&=& \frac{1}{Z}\sum_n \e^{-(\beta+it)E_n^0} \langle \Psi^0_n|\e^{iH_1t}|\Psi^0_n\rangle \nonumber \, .
\end{eqnarray}
The result is a thermal average over the Loschmidt echo of pure states
and thus very different from the overlap between density matrices
defined in Eq.~\eqref{LoT}.

For all generalized Loschmidt echos discussed here an appropriate
return rate \eqref{return} can be defined. It is the return rate in
the thermodynamic limit which we want to study in the following.

\section{Unitary dynamics in Gaussian models}
\label{Gaussian}

We consider free fermion models described by the Hamiltonian
\begin{equation}
\label{Gmodel}
H=\sum_{k\ge 0} \Psi_k^\dagger \mathcal{H}_k \Psi_k
\end{equation}
with $\Psi_k=(c_k,c_{-k}^\dagger)^T$. Here $c_k$ is an annihilation
operator of spinless fermions with momentum $k$. This Hamiltonian
describes models with a single-site unit cell which are bilinear in
the creation and annihilation operators and can contain pairing terms
as in the transverse Ising and Kitaev chains, see Sec.~\ref{Sec_Ising}. If
we identify $d_k\equiv c_{-k}^\dagger$ then the Hamiltonian
\eqref{Gmodel} can also describe models with a two-site unit cell
which contain only hopping and no pairing terms such as the SSH and
Rice-Mele models, see Sec.~\ref{Sec_SSH}. The momentum summation in
both cases runs over the first Brillouin zone. It is often convenient
to write the $2\times 2$ matrix $\mathcal{H}_k$ as $\mathcal{H}_k
=\mathbf{d}_k\cdot\mathbf{\sigma}$ where $\mathbf{d}_k$ is a
three-component parameter vector and $\mathbf{\sigma}$ the vector of
Pauli matrices. During the quench the parameter vector $\mathbf{d}_k$
is changed leading to an initial Hamiltonian $H_0$ and a final
Hamiltonian $H_1$. In the two different bases in which the
Hamiltonians are diagonal we have
\begin{equation}
\label{Gmodel2}
H_{i}=\sum_{k\ge 0} \varepsilon_k^{i}\left( c_{ki}^\dagger c_{ki} +c_{-ki}^\dagger c_{-ki}-1 \right)
\end{equation}
with energies $\varepsilon_k^i>0$ and $i=0,1$. The operators in which
the two Hamiltonians are diagonal are related by a Bogoliubov
transform
\begin{equation}
\label{Bogo}
c_{k0} = u_k c_{k1} + v_k c_{-k1}^\dagger \; ;\; c_{k1} = u_k c_{k0} -v_k c_{-k0}^\dagger.
\end{equation}
The Bogoliubov variables can be parametrized by an angle $\theta_k$
as $u_k=\cos\theta_k$ and $v_k=\sin\theta_k$. For each $k$-mode there
are $4$ basis states. We can either work in the eigenbasis
$|\Psi_j^0\rangle$ of $H_0$ or the eigenbasis $|\Psi_j^1\rangle$ of
$H_1$ which can be expressed as
\begin{eqnarray}
\label{trafo}
|\Psi_0^0\rangle &=& |0\rangle_0 =(u_k - v_k c_{k1}^\dagger c_{-k1}^\dagger) |0\rangle_1 \nonumber \\
|\Psi_1^0\rangle &=& c_{k0}^\dagger |0\rangle_0 = c_{k1}^\dagger |0\rangle_1 \nonumber \\
|\Psi_2^0\rangle &=& c_{-k0}^\dagger |0\rangle_0 = c_{-k1}^\dagger |0\rangle_1 \\
|\Psi_3^0\rangle &=& c_{k0}^\dagger c_{-k0}^\dagger |0\rangle_0 = (v_k + u_k c_{k1}^\dagger c_{-k1}^\dagger) |0\rangle_1 \nonumber  
\end{eqnarray}
or vice versa
\begin{eqnarray}
\label{trafo2}
|\Psi_0^1\rangle &=& |0\rangle_1 =(u_k + v_k c_{k0}^\dagger c_{-k0}^\dagger) |0\rangle_0 \nonumber \\
|\Psi_1^1\rangle &=& c_{k1}^\dagger |0\rangle_1 = c_{k0}^\dagger |0\rangle_0 \nonumber \\
|\Psi_2^1\rangle &=& c_{-k1}^\dagger |0\rangle_1 = c_{-k0}^\dagger |0\rangle_0 \\
|\Psi_3^1\rangle &=& c_{k1}^\dagger c_{-k1}^\dagger |0\rangle_1 = (-v_k + u_k c_{k0}^\dagger c_{-k0}^\dagger) |0\rangle_0 \nonumber \, . 
\end{eqnarray}
Here $|0\rangle_{0,1}$ are the ground states of $H_{0,1}$. The
Loschmidt echo at zero temperature can be easily calculated using the
transformation \eqref{trafo} leading to
\begin{eqnarray}
\label{LT0}
{\cal L}_0(t) &= &\prod_k \left[ u_k^2 \e^{i\varepsilon_k^1 t} + v_k^2 \e^{-i\varepsilon_k^1 t}\right] \\
&=& \prod_k \left[ \cos\left(\varepsilon_k^1 t\right)+ i\sin(2\theta_k)\sin\left(\varepsilon_k^1 t\right)\right] \nonumber
\end{eqnarray}
and $|{\cal L}_0(t)|^2 = \prod_k |{\cal L}_0^k(t)|^2$ with
\begin{equation}
\label{LT0p}
|{\cal L}_0^k(t)|^2= \left[ 1-\sin^2(2\theta_k)\sin^2\left(\varepsilon_k^1 t\right)\right] \, .
\end{equation}
Here $\cos(2\theta_k)=\hat{\mathbf{d}}_k^0\cdot\hat{\mathbf{d}}_k^1$
with $\hat{\mathbf{d}}^i_k$ being the normalized parameter vector. Note
that the result \eqref{LT0} is also valid for free fermion models with
a two-site unit cell but without pairing terms although the ground
state is different. From \eqref{LT0p} it is evident that ${\cal L}_0(t_c)=0$
if a momentum $k_c$ exists with
$\hat{\mathbf{d}}_{k_c}^0\cdot\hat{\mathbf{d}}_{k_c}^1=0$, i.e. $\sin(2\theta_k)=1$. The
critical times are then given by
\begin{equation}
\label{tc}
t_c=\frac{\pi}{2\varepsilon_{k_c}^1}(2n+1).
\end{equation}
For any of the generalized Loschmidt echos defined before we can write
the return rate as
\begin{equation}
\label{Gaussian_return}
l(t) =-\frac{1}{2\pi}\int\ln|L^k(t)|\, dk \, .
\end{equation}
In the following we will explicitly calculate $l(t)$ for the different generalized Loschmidt echos. 

\hspace*{0.2cm}

\subsection{Projection onto a pure state}
We want to first investigate the case where only one of the states
is impure. A natural generalization is then the Loschmidt echo
defined in Eq.~\eqref{Lproj}. For the considered Gaussian models
\eqref{Gmodel} the Loschmidt echo separates into a product $|{\cal L}_p(t)|^2=\prod_k
|{\cal L}_p^{k}(t)|^2$. If we, furthermore, assume that our initial
mixed state is described by a canonical ensemble then we obtain
\begin{eqnarray}
\label{Lproj2}
|{\cal L}_p^{k}(t)|^2 &=& {\langle \Psi_0^0|\rho_k(t)|\Psi_0^0\rangle}/{\langle\Psi_0^0|\rho_k(0)|\Psi_0^0\rangle} \\
&=& \sum_{n=0}^3 \e^{-\beta (E_{kn}^0-E_{k0}^0)} |\langle \Psi_0^0|\e^{-iH_1t}|\Psi_n^0\rangle|^2 \nonumber
\end{eqnarray}
where we have used the spectral representation of the density matrix
$\rho_k(t)$ in terms of the eigenstates of $\mathcal{H}_k^0$ and
$\beta$ is the inverse temperature. The eigenenergies of the $4$
eigenstates for each $k$-mode are denoted by $E_{kn}^0=\bigl(-\varepsilon_k^0,0,0,\varepsilon_k^0\bigr)$. Using the
representation
\eqref{trafo} of the eigenstates in terms of the operators of the
final Hamiltonian $H_1$ one finds
\begin{equation}
\label{Lproj3}
|{\cal L}_p(t)|^2=\prod_k \left[
1-\left(1-\e^{-2\beta\varepsilon_k^0}\right)\sin^2(2\theta_k)\sin^2(\varepsilon_k^1
t)\right].
\end{equation}
It is obvious that ${\cal L}_p(t)=0$ is only possible at zero temperature in
which case $|{\cal L}_p(t)|\equiv |{\cal L}_0(t)|$, see Eq.~\eqref{LT0p}. If one starts
from a mixed state then the DPT's are washed out even if one projects
onto the ground state. With the appropriately chosen ground state and
the associated energies $E_{kn}^0$, the result
\eqref{Lproj3} also holds for the models with a two-site unit cell
such as the SSH and Rice-Mele models.

\subsection{Thermal density matrices}
\label{Thermal}
The calculation of \eqref{LoT} for the case that $\rho(0)$ is a
thermal density matrix is instructive for the dissipative case
discussed in Sec.~\ref{Open} so we briefly rederive the known
result\cite{Zanardi2007a,CamposVenuti2011} for ${\cal L}_\rho(t)$
here. It is most convenient to perform the calculation in the
eigenbasis of the time-evolving Hamiltonian $H_1$ using the
transformation \eqref{trafo2}. Because only the states
$|\Psi_0^i\rangle$ and $|\Psi_3^i\rangle$ are mixed by the
transformation, the initial unnormalized density matrix $\rho_k(0)$
can be rearranged into two $2\times 2$ block matrices $\mathbf{I}_2$
(identity matrix) and $\mathbf{r}_k(0)$ with
\begin{widetext}
\begin{equation}
\label{r_k}
{\bm r}_{k}(0)=
\begin{pmatrix}
\cosh\left(\beta\varepsilon^0_k\right)+\sinh\left(\beta\varepsilon^0_k\right)\cos(2\theta_k) &
-\sinh\left(\beta\varepsilon^0_k\right)\sin(2\theta_k)\\
-\sinh\left(\beta\varepsilon^0_k\right)\sin(2\theta_k) & 
\cosh\left(\beta\varepsilon^0_k\right)-\sinh\left(\beta\varepsilon^0_k\right)\cos(2\theta_k) 
\end{pmatrix}\,.
\end{equation}
\end{widetext}
$\sqrt{r_k(0)}$ is obtained from \eqref{r_k} by replacing
$\beta\to\beta/2$ and $r_k(t)$ by replacing $r_k^{(12)}\to
\e^{2i\varepsilon_k^1 t}r_k^{(12)}$ and $r_k^{(21)}\to
\e^{-2i\varepsilon_k^1 t}r_k^{(21)}$. The partition function is 
given by
$Z_k=\Tr\rho_k=\Tr(\mathbf{I}_2)+\Tr\mathbf{r}_k(0)=2+2\cosh(\beta\varepsilon_k^0)$. We
can now simplify the generalized Loschmidt echo \eqref{LoT} in this case to
\begin{equation}
\label{LoT2}
{\cal L}_\rho(t)=\prod_k\frac{2+\lambda_{k1}(t)+\lambda_{k2}(t)}{2+2\cosh(\beta\varepsilon_k^0)}
\end{equation}
where $\lambda_{ki}^2(t)$ are the eigenvalues of
$\sqrt{\mathbf{r}_k(0)}\mathbf{r}_k(t)\sqrt{\mathbf{r}_k(0)}$
which are given by
\begin{equation}
\label{l12}
\lambda_{k1,2}(t)=\sqrt{1+|{\cal L}_0^k(t)|^2\sinh^2[\beta\epsilon^0_k]}\pm |{\cal L}_0^k(t)|\sinh[\beta\epsilon^0_k]\,,
\end{equation}
with ${\cal L}_0^k(t)$ defined in Eq.~\eqref{LT0p}. As a final
result we thus obtain\cite{Zanardi2007a,CamposVenuti2011}
\begin{equation}
\label{LoT3}
{\cal L}_\rho(t)=\prod_k\frac{1+\sqrt{1+|{\cal L}_0^k(t)|^2\sinh^2(\beta\varepsilon^0_k)}}{1+\cosh(\beta\varepsilon^0_k)}\,.
\end{equation}
For any finite temperature this means that ${\cal L}_\rho(t)>0$ for all
times, i.e., there are no DPT's. For $\beta\to\infty$ the result
reduces to the zero-temperature result, Eq.~\eqref{LT0p}. The result
\eqref{LoT3} also holds for Gaussian models with a two-site unit cell
such as the SSH and Rice-Mele models.

We now also briefly discuss the possible generalization $\tilde
{\cal L}_\rho(t)$ defined in Eq.~\eqref{Ltilde}. While this function, in
general, does not fulfill the requirements listed in Sec.~\ref{DM_Lo}
it turns out that for the case considered here at least $0\leq |\tilde
{\cal L}_\rho(t)|\leq 1$ is fulfilled. We start again from a thermal density
matrix. The spectral representation using the eigenstates of $H_1$
then reads
\begin{equation}
\label{Ltilde2}
|\tilde {\cal L}_\rho(t)|^2=\frac{\sum_{n,m}\e^{i(E_m^1-E_n^1)t}|\langle\Psi_n^1|\e^{-\beta H_0}|\Psi_m^1\rangle|^2}{\sum_n \e^{-2\beta E_n^0}}\, .
\end{equation}
Only the eigenstates $|\Psi_0^1\rangle$ and $|\Psi_0^3\rangle$ mix and
it is easy to check the final result
\begin{eqnarray}
\label{Ltilde3}
&&|\tilde {\cal L}_\rho(t)|^2 = \prod_k\left[ \cosh^{-2}(\beta \varepsilon_k^0) +\tanh^2(\beta\varepsilon_k^0) |{\cal L}_0^k(t)|^2\right] \nonumber \\
&& = \prod_k\left[1-\tanh^2(\beta\varepsilon_k^0) \sin^2(2\theta_k)\sin^2(\varepsilon_k^1 t)\right] \, .
\end{eqnarray}
$\tilde {\cal L}_\rho(t)=0$ is again only possible if $T=0$.

\subsubsection{Ising and Kitaev models}
\label{Sec_Ising}
\begin{figure}
\includegraphics[width=0.99\columnwidth]{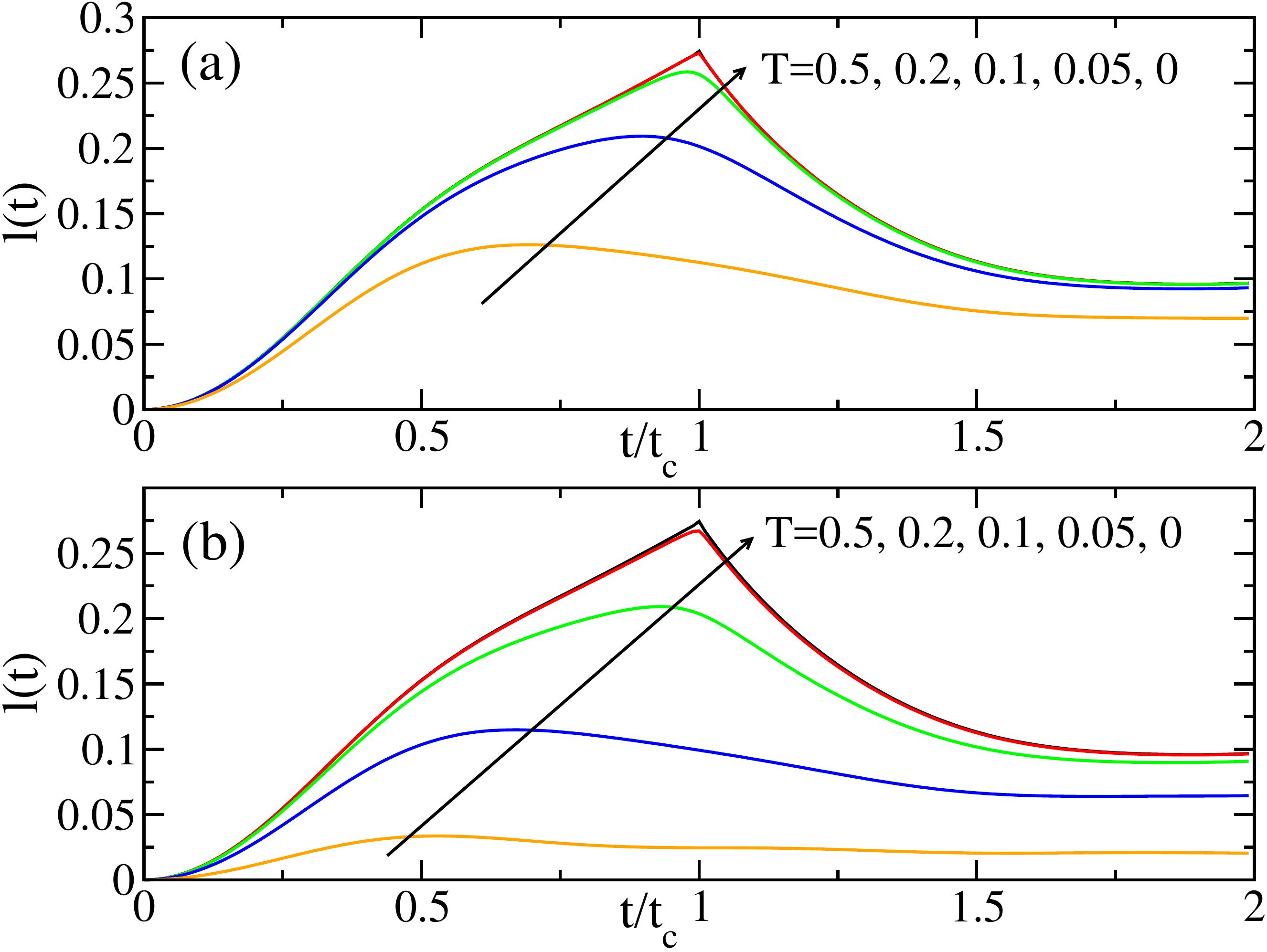}
\caption{(Color online) The return rate $l(t)$ for the Ising chain in the thermodynamic limit 
for a quench from $g=0.5$ to $g=1.5$ at different temperatures
$T$. (a) Projection onto the ground state, Eq.~\eqref{Lproj3} (note
that the curves for $T=0$ and $T=0.05$ are almost on top of each
other), and (b) generalized Loschmidt echo, Eqs.~\eqref{LoT}, and \eqref{LoT3}.}
\label{Fig1}
\end{figure}

The finite-temperature results can be directly applied to concrete
models. The Kitaev chain, for example, is defined by
\begin{equation}
\label{kh}
H=\sum_{i}\left[\Psi^\dagger_{i}\left(
\Delta\im{\bm\tau}^y-J{\bm\tau}^z\right)\Psi_{i+1}+\textrm{H.c.}-\Psi^\dagger_{i}\mu{\bm\tau}^z\Psi_{i}\right]\,
\end{equation}
where $\Psi^\dagger_{i}=(c^\dagger_{i},c_{i})$ and $c_{
i}^{(\dagger)}$ annihilates (creates) a spinless particle at site
$i$. The Kitaev chain is topologically non-trivial when $\mu<2|J|$ and
$\Delta\neq0$. Note that $\Delta=0$ is a phase boundary between
phases with winding numbers $\pm1$. As a special case the transverse
Ising model
\begin{equation}
\label{Ising}
H(g)=-\frac{1}{2}\sum_{i}{\bm\sigma}^z_i{\bm\sigma}^z_{i+1}+\frac{g}{2}\sum_{i=1}^N{\bm\sigma}^x_i
\end{equation}
is obtained if one sets $\mu=-g/2$ and $J=1/4=-\Delta$ in
\eqref{kh}. After a Fourier transform, for a chain with periodic
boundary conditions, the Hamiltonian \eqref{kh} is of the form of
Eq.~\eqref{Gmodel} with parameter vector
\begin{equation}
\mathbf{d}_k=\begin{pmatrix}
0,2\Delta\sin k,-2J\cos k -\mu
\end{pmatrix}\,,
\end{equation}
and
$\cos(2\theta_k)=\hat{\mathbf{d}}_k^0\cdot\hat{\mathbf{d}}_k^1$. In
Fig.~\ref{Fig1} we plot the return rate in the thermodynamic limit,
Eq.~\eqref{Gaussian_return}, for a quench from $g=0.5$ to $g=1.5$.
While the cusp in the return rate at the critical time $t_c$ is only
slightly rounded off for temperatures up to $T=0.1$ if we project onto
the ground state, Eq.~\eqref{Lproj}, signatures of a DPT are
already almost lost at this temperature if we use the generalized
Loschmidt echo \eqref{LoT} which measures the distance between the
initial and the time-evolved thermal density matrix. 

\subsubsection{SSH and Rice-Mele models}
\label{Sec_SSH}
The Rice-Mele and the SSH chains are models with a two-site unit cell
and alternating hoppings $1\pm\delta$ and potentials $\pm V$. The
Hamiltonian for the Rice-Mele model is given by
\begin{eqnarray}
\label{RM_model}
H&=&\sum_i\Psi^\dagger_i\left[-(1+\delta){\bm \sigma}^x+V{\bm \sigma}^z\right]\Psi_i\\\nonumber&&-(1-\delta)\sum_j\Psi^\dagger_i
\begin{pmatrix}0&0\\1&0\end{pmatrix}\Psi_{i+1}+\textrm{H.c.}
\end{eqnarray}
with $\Psi_i=(c_i,d_i)$. After a Fourier transform this model can also
be represented by the generic Hamiltonian \eqref{Gmodel} with the
identification $d_k\equiv c_{-k}^\dagger$. The parameter vector in this
case is given by
\begin{equation}
\mathbf{d}_k=\begin{pmatrix}
-2 \cos k, 2\delta\sin k,V
\end{pmatrix}\,.
\end{equation}
The SSH model is a special case of the Rice-Mele model obtained by
setting the alternating potential $V=0$.

\begin{figure}
\includegraphics[width=0.99\columnwidth]{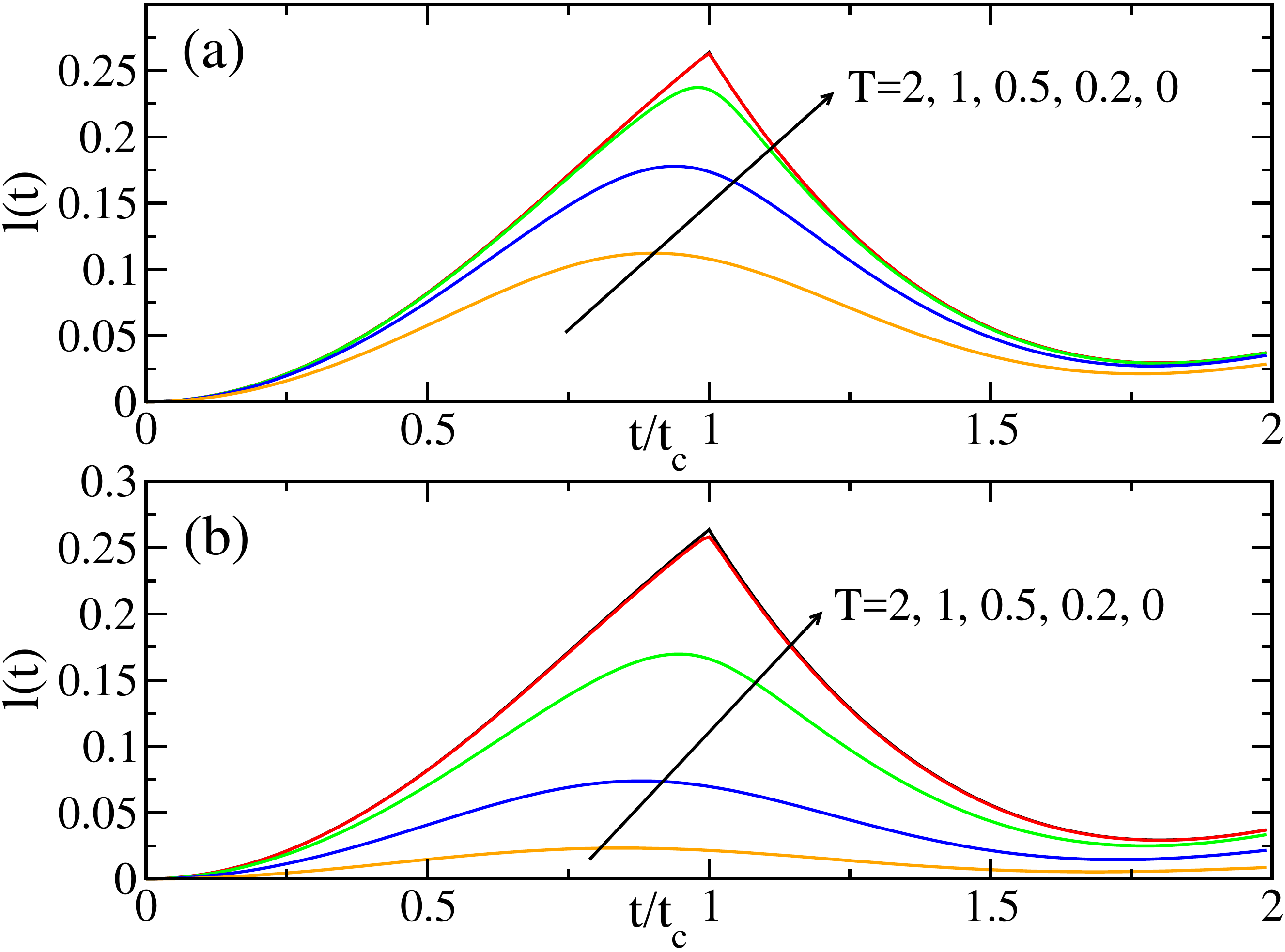}
\caption{(Color online) The return rate $l(t)$ for the SSH chain in the thermodynamic limit 
for a quench from $\delta=-0.5$ to $\delta=0.5$ at different
temperatures $T$. (a) Projection onto the ground state,
Eq.~\eqref{Lproj3}, and (b) generalized Loschmidt echo,
Eq.~\eqref{LoT3}. Note that the curves for $T=0$ and $T=0.2$ are
almost on top of each other.}
\label{Fig2}
\end{figure}

In Fig.~\ref{Fig2} the return rate for a symmetric quench from
$\delta=-0.5$ to $\delta=0.5$ for $V=0$ is shown.

While the cusp in the return rate at the critical time $t_c$ is washed
out in this case as well, a signature of the DPT at zero temperature
is more cleary visible also at finite temperatures as compared to the
quench in the Ising model shown in Fig.~\ref{Fig1}.

\section{Open systems}
\label{Open}
In systems where the Loschmidt echo has been studied experimentally
such as cold atomic gases and trapped
ions\cite{Flaschner2016,Jurcevic2017} interactions with
electromagnetic fields are used to control the particles. These
systems are therefore intrinsically open systems and decoherence and
loss processes are unavoidable. Using the Born-Markov approximation
such open systems can be described by a Lindblad master equation
\begin{equation}
\label{LME}
\dot \rho(t) = -i [H,\rho] + \sum_\mu \left(L_\mu\rho L_\mu^\dagger -\frac{1}{2}\left\{L_\mu^\dagger L_\mu,\rho\right\}\right).
\end{equation}
Here $L_\mu$ are the Lindblad operators describing the dissipative,
non-unitary dynamics induced by independent reservoirs labelled by $\mu$,
and $\{\cdot,\cdot\}$ is the anti-commutator. In
order to have a bilinear LME which can be solved exactly, we continue
to consider Hamiltonians as defined in Eq.~\eqref{Gmodel} with
periodic boundary conditions which can be diagonalized in Fourier
space. 
We consider Lindblad operators that are linear in creation and annihilation operators
leading to a linear dynamics
\begin{equation}
\label{Lindblad1}
L_\mu= \sqrt{\gamma_{\mu}} c_{\mu} \; \mbox {and} \; L_\mu= \sqrt{\bar\gamma_{\mu}} c^\dagger_{\mu}
\end{equation}
describing particle loss and creation processes with amplitudes
$\gamma_{\mu}>0$ and $\bar\gamma_{\mu}>0$, respectively.  This form
ensures that the dissipative terms in Eq.~\eqref{LME} are also
bilinear.
More specifically we consider reservoirs that couple each to only one $k$-mode
\begin{equation}
\label{Lindblad2}
L_k= \sqrt{\gamma_{\pm k}} c_{\pm k} \; \mbox {and} \; L_k= \sqrt{\bar\gamma_{\pm k}} c^\dagger_{\pm k} \; .
\end{equation}

To solve the Lindblad equation we will use the superoperator
formalism.\cite{Carmichael1998} The $n\times n$ density matrix $\rho$ is
recast into an $n^2$-dimensional vector $||\rho\rangle\rangle$ and the
Hamiltonian and Lindblad operators become superoperators acting on this
vector. The LME \eqref{LME} and its solution can then be written as
\begin{equation}
\label{LME2}
||\dot\rho\rangle\rangle =\mathcal{L}\, ||\rho\rangle\rangle \quad ; \quad ||\rho\rangle\rangle(t)\, =\exp(\mathcal{L}t)\,||\rho(0)\rangle\rangle \, .
\end{equation}
For the purely unitary time evolution considered in the previous
section the Lindbladian $\mathcal{L}$ takes the form
\begin{equation}
\label{HLind}
\mathcal{L} = -i\left(H\otimes\mathbf{I}_n + \mathbf{I}_n\otimes H^\dagger\right)
\end{equation}
where $\mathbf{I}_n$ is the $n\times n$ identity matrix. Similarly, the
individual Lindblad operators \eqref{Lindblad2} can be
written as superoperators acting on $||\rho\rangle\rangle$. The
solution vector $||\rho\rangle\rangle(t)$ can then be recast into a
matrix allowing one to calculate the generalized Loschmidt echos also for
open systems.

\subsection{Particle loss}

We consider again free fermionic models of the type
\eqref{Gmodel} with the $4$ basis states \eqref{trafo} for each
$k$-mode. 

As a first example, we investigate a simple mixed initial state
$\rho_k(0)=\frac{1}{2}\left(|\Psi_1^0\rangle\langle\Psi_1^0| +
|\Psi_2^0\rangle\langle\Psi_2^0|\right)$ and a time evolution under
the Lindblad operators $L_{1k}=\sqrt{\gamma_k}c_k$ and
$L_{2k}=\sqrt{\gamma_{-k}}c_{-k}$. In this case it is straightforward
to show that the density matrix takes the form
$\rho_k(t)=\frac{1}{2}\text{diag}(2-\e^{-\gamma_kt}-\e^{-\gamma_{-k}t},\e^{-\gamma_kt},\e^{-\gamma_{-k}t},0)$. The
non-equilibrium steady state (NESS) is thus the completely empty state
for $\gamma_{\pm k}\neq 0$. Since both $\rho(0)$ and $\rho(t)$ are
diagonal it follows immediately that the generalized Loschmidt echo is
given by
\begin{equation}
\label{Lrholoss}
{\cal L}_\rho(t)=\frac{1}{2}\prod_k\left(\e^{-\gamma_k t/2} + \e^{-\gamma_{-k}t/2}\right) \, .
\end{equation}
As one might have expected, ${\cal L}_\rho(t)$ shows an exponential decay in
this case. If $\gamma_k=\gamma_{-k}=\gamma=\mbox{const}$ then the
return rate in the thermodynamic limit \eqref{Gaussian_return}
increases linearly, $l(t)=\gamma t/2$, and thus diverges only at
infinite time.

\subsection{Quench in Kitaev-type models with particle loss}
\label{Kitaevloss}

Next, we want to consider a quench for a Kitaev-type model with
Hamiltonian \eqref{Gmodel2} with the basis states \eqref{trafo2}. As
in Sec.~\ref{Thermal} we start with a thermal density matrix $\rho(0)$
but now also allow for particle loss processes as in the example
above. Crucially, the matrix $\rho_k(t)$ still can be decomposed into
two $2\times 2$ block matrices. We can therefore write
${\cal L}_\rho^k(t)=\Tr\sqrt{M_1}+\Tr\sqrt{M_2}$ with
$M_i=\sqrt{\rho^i_k(0)}\rho^i_k(t)\sqrt{\rho^i_k(0)}$ and $\rho_k^{1,2}$
being the two block matrices. With
$\Tr\sqrt{M_i}=\sqrt{\lambda^i_1}+\sqrt{\lambda^i_2}>0$ we can write
$\left(\Tr\sqrt{M_i}\right)^2=\lambda^i_1+\lambda^i_2+2\sqrt{\lambda^i_1\lambda^i_2}=\Tr
M_i+2\sqrt{\det M_1}$.\cite{Jozsa1994} For the Loschmidt echo we
therefore find
\begin{equation}
\label{LoTloss}
{\cal L}_\rho(t)=\prod_k\sum_{i=1,2}\sqrt{\Tr M_i +2\sqrt{\det M_i}} \, .
\end{equation}
Using this formula it is straightforward to obtain an explicit result
for ${\cal L}_\rho(t)$ which, however, is quite lengthy for finite
temperatures. We therefore limit ourselves here to presenting the
result for $T=0$ only. In this case one of the block matrices is zero
and we obtain the following closed-form expression
\begin{eqnarray}
\label{LoTloss2}
L^2_\rho(t)&=&\prod_k \e^{-\Gamma^+_k t}\bigg[
\cos 2\theta_k\sinh\left(\Gamma_k^+ t\right) -\sin^2 2\theta_k \sin^2(\varepsilon_k^1 t)\nonumber \\
&+& 
\left.\frac{1}{2}\sin^22\theta_k\left(1-\cosh(\Gamma_k^- t)\right)+\cosh(\Gamma_k^+ t)
\right] \, .
\end{eqnarray}
Here we have defined $\Gamma_k^\pm =(\gamma_k \pm\gamma_{-k})/2$. It is
easy to see that this result reduces to Eq.~\eqref{LT0p} for
$\gamma_k=\gamma_{-k}=0$. Furthermore, there are no DPT's for finite
loss rates. 

As an example for the broadening of the cusps in the return rate
\eqref{Gaussian_return} we consider the same quench in the transverse
Ising model as before.
\begin{figure}
\includegraphics[width=0.99\columnwidth]{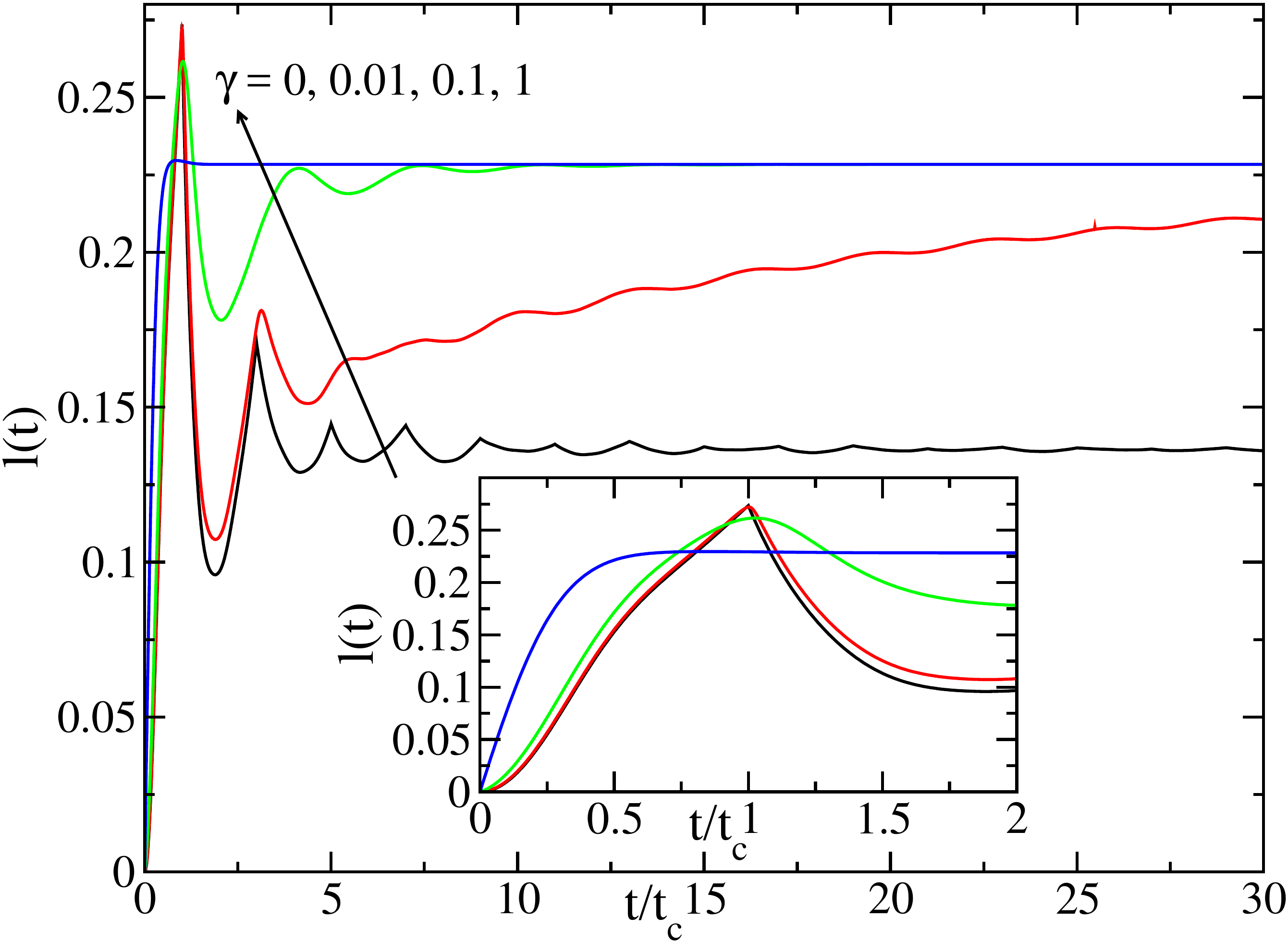}
\caption{(Color online) The return rate $l(t)$ for the Ising chain in the thermodynamic limit for a quench from $g=0.5$ to $g=1.5$ at $T=0$ for different particle loss rates $\gamma=\gamma_k=\gamma_{-k}$. Inset: Broadening of the first cusp at $t=t_c$.}
\label{Fig3}
\end{figure}
Fig.~\ref{Fig3} shows that small loss rates already lead to a
significant broadening of the first cusp at $t=t_c$ and completely
wash out the cusps at longer times. Furthermore, the NESS for a
non-zero loss rate is always the empty state so that the return rate
at infinite times becomes {\it independent} of the loss rate and is
given by
\begin{equation}
\label{returntinfty}
l(t\to\infty)=-\frac{1}{2\pi}\int_0^\pi\ln\left(\frac{1+\hat{\mathbf{d}}_k^0\cdot\hat{\mathbf{d}}_k^1}{2}\right)\, dk \, .
\end{equation}

\subsection{Quench in Kitaev-type models with particle creation and loss}
\label{Kitaevcreation}

So far we have seen that both finite temperatures and particle loss
processes destroy DPT's. One can then ask if it is possible to
engineer dissipative processes in an open quantum system in such a way
that DPT's persist. By constructing a concrete example we will show
that this is indeed possible. 

We consider the case that particles with momentum $k$ are annihilated
with rate $\gamma_k$ while particles with momentum $-k$ are created
with rate $\bar\gamma_{-k}$. As in the case with particle loss
considered in Sec.~\ref{Kitaevloss} the density matrix $\rho_k(t)$
still has block structure and a calculation along the same lines is
possible. At $T=0$ we obtain a result which is very similar to
Eq.~\eqref{LoTloss2} and reads
\begin{eqnarray}
\label{LoTcreation}
L^2_\rho(t)&=&\prod_k \e^{-\tilde\Gamma^+_k t}\bigg[ \cos 2\theta_k\sinh(\tilde\Gamma_k^- t) -\sin^2 2\theta_k \sin^2(\varepsilon_k^1 t) \nonumber \\
&+&\left.\frac{1}{2}\sin^22\theta_k(1-\cosh(\tilde\Gamma_k^- t))+\cosh(\tilde\Gamma_k^- t) 
\right].
\end{eqnarray}
The rates are now defined as $\tilde\Gamma_k^\pm =(\gamma_k
\pm\bar\gamma_{-k})/2$. The essential difference when comparing
Eq.~\eqref{LoTcreation} with the previous result \eqref{LoTloss2} is
that inside the bracket only the rate $\tilde\Gamma_k^-$ is
present. For $\tilde\Gamma_k^-=0$, i.e. $\gamma_k=\bar\gamma_{-k}$,
the Loschmidt echo becomes $L^2_\rho(t)=\prod_k\exp(-\tilde\Gamma^+_k
t)|{\cal L}_0^k(t)|^2$ which is the zero-temperature result \eqref{LT0p} with an
additional exponential decay. DPT's are thus still present for this
particular case at the same critical times $t_c$ despite the
dissipative processes.

\begin{figure}[htp]
\includegraphics[width=0.99\columnwidth]{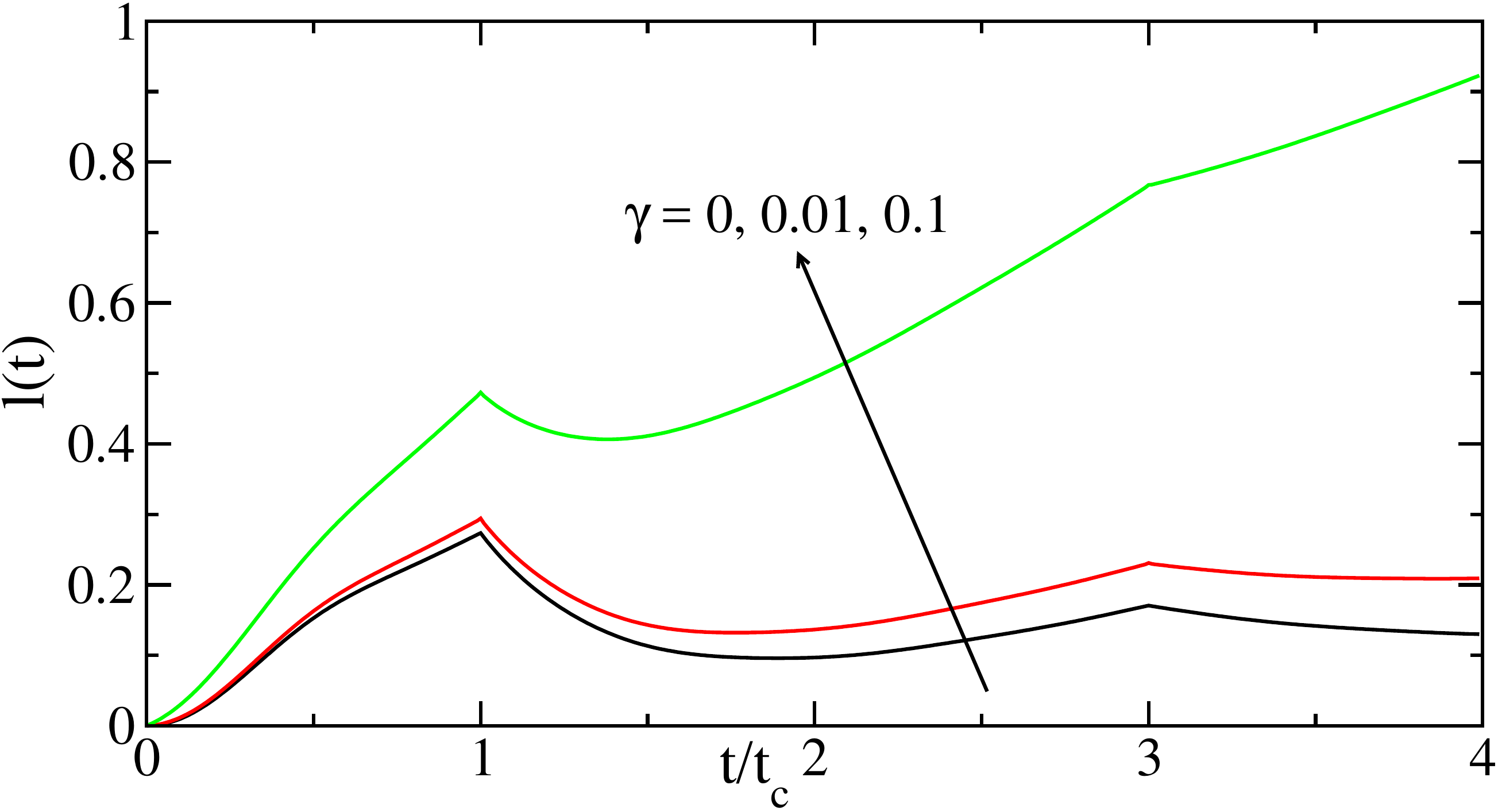}
\caption{(Color online) The return rate $l(t)$ for the Ising chain in the 
thermodynamic limit for a quench from $g=0.5$ to $g=1.5$ at $T=0$ for
various equal particle loss and creation rates
$\gamma=\gamma_k=\bar\gamma_{-k}$.}
\label{Fig4}
\end{figure}

As an example, we consider again the quench in the transverse Ising
chain. In Fig.~\ref{Fig4} we show results for the fine-tuned point
$\gamma=\gamma_k=\bar\gamma_{-k}$.
The cusps remain clearly visible for finite dissipation rates. For a
$k$ independent rate $\tilde\Gamma_k^+\equiv \tilde\Gamma^+$ as chosen
in Fig.~\ref{Fig4} the result for the return rate is
\begin{equation}
\label{Ising_return_creation}
l(t)=\frac{\tilde\Gamma^+ t}{2}-\frac{1}{\pi}\int_0^\pi \ln |{\cal L}_0^k(t)| \, dk \, .
\end{equation}
This is simply the zero-temperature return rate in the closed system
plus a linear increase with slope $\tilde\Gamma^+/2$. In the NESS at long
times all particles will be in the $-k$ states leading to a vanishing
Loschmidt echo and a diverging return rate.

\section{Conclusions}
\label{Concl}

We have studied a generalization of the Loschmidt echo to density
matrices which is applicable both to finite temperatures and to open
systems. It is based on a direct generalization of the fidelity for
mixed states to dynamical problems and provides a measure of the
distance between the initial and the time-evolved density matrix. As
such it is very different from previous generalizations studied in the
context of dynamical phase transitions which are based on thermal
averages over the Loschmidt echos of pure states and are only
applicable to unitary dynamics.

For bilinear one-dimensional fermionic lattice models with periodic
boundary conditions we have shown that finite temperatures always wash
out the non-analyticities in the return rate of the generalized
Loschmidt echo. Dynamical phase transitions only exist at zero
temperature.

For open quantum systems described by a Lindblad master equation we
similarly find that particle loss processes smooth out cusps in the
return rate so that signatures of the dynamical phase transition are
hard to detect even if the loss rates are very small.

Finally, we showed that it is possible to fine-tune particle loss and
creation processes in such a way that dynamical phase transitions can
be observed despite the dissipative dynamics.

The generalized Loschmidt considered in this paper can be understood
as a tool to measure distances between density matrices. As such it
might be helpful in engineering and controlling specific states using
dissipative dynamics. Zeros of the Loschmidt echo signal, in
particular, that a mixed state has been reached such that all
purifications to states in an enlarged Hilbert space are orthogonal to
purifications of the initial state.

\hspace*{0.4cm}

Shortly after submitting this paper Ref.~\onlinecite{Mera2017} became
available, which is on a related topic.

\acknowledgments
JS acknowledges support by the Natural Sciences and Engineering
Research Council (NSERC, Canada) and by the Deutsche
Forschungsgemeinschaft (DFG) via Research Unit FOR 2316. 
MF acknowledges support by the Deutsche
Forschungsgemeinschaft (DFG) via the Collaborative Research Center SFB-TR 185.

%


\end{document}